\documentstyle[12pt,epsf]{article}
\def\reference{\parskip 0pt\par\noindent\hangindent 0.5 truecm}
\def\kms{km ${\rm s}^{-1}$}
\newcommand{\hi}{H{\sc i} }
\newcommand{\mhi}{M_{\rm HI}}
\newcommand{\msol}{{M}_\odot}
\newcommand{\ohi}{\Omega_{\rm HI}}
\newcommand{\rhi}{\rho_{\rm HI}}
\begin{document}
%
%
\title{An \hi selected sample of galaxies --- The \hi mass function
and the surface brightness distribution}
\author{Martin Zwaan,
        Frank Briggs and
        David Sprayberry\\Kapteyn Astronomical Institute\\
	  P.O. Box 800, NL-9700 AV Groningen, The
Netherlands\\zwaan@astro.rug.nl, fbriggs@astro.rug.nl, dspray@astro.rug.nl}
\date{} 
\maketitle

\begin{abstract}
 Results from the Arecibo \hi Strip Survey, an unbiased extragalactic
\hi survey, combined with optical and 21cm follow-up observations,
determine the \hi Mass Function and the cosmological mass density of
\hi at the present epoch.  Both are consistent with earlier estimates,
computed for the population of optically selected galaxies.  This
consistency occurs because, although the distribution of optical central
surface brightnesses among galaxies is flat, we fail to find a
population of galaxies with central surface brightnesses fainter than
24~$B$-$\rm mag\, arcsec^{-2}$, even though there is no
observational selection against them. 
 \end{abstract}

\bigskip

\section{Introduction}
 There have been speculations that low surface brightness (LSB) galaxies
and intergalactic clouds might constitute a substantial portion of the
population of nearby extragalactic objects and that they might contain
comparable mass to that in normal galaxies.  The LSB population would
escape detection optically and would not be included in the galaxy
luminosity functions that are used to compute the visible baryonic
content of the local Universe (Disney 1976, McGaugh 1996).  On the other
hand, estimates of the \hi mass function (HIMF) based on published
observations (Briggs 1990) have seemed to indicate that there is
probably not any substantial population that has been missed.  Weinberg
et al (1991) and Szomoru et al (1994) have come to the same conclusion. 
Until recently, there were no galaxy samples that could be used to
address this question empirically, since the galaxies were all first
identified optically. 

In this paper we present results from the Arecibo \hi Strip Survey, an
unbiased 21cm survey with adequate sensitivity to detect \hi of very low
surface density.  It is of sufficient length (approximately 15 hours of
RA) and depth (7400 \kms) that it should be immune to fluctuations due
to the large scale structure.  The total sky coverage was $\sim\!65$
square degrees.  In the main beam, which has a FWHM of 3.2~arcmin, the
survey was capable of detecting \hi masses of $6\times 10^5 h^{-2}
\msol$ at 7 $h^{-1}$ Mpc and $1.5\times 10^8 h^{-2} \msol$ at the full
depth of the survey.  The details of the Arecibo Strip Survey are
described by Sorar (1994) and Briggs (1996). 

The survey yielded a total of 61 detections, of which about half could
be associated with cataloged galaxies listed in the NASA Extragalactic
Database (NED).  About five detections with galactic latitude
$|b|>10^{\circ}$, where extinction is not a problem, have no obvious
counterparts on the Digitized Sky Survey (DSS).  The \hi selected
galaxies generally follow the same structures as optical selected
galaxies.  This is consistent with (1) results from Szomoru et al (1996)
who fail to detect large numbers of \hi selected galaxies in selected
void fields and (2) the finding that LSB galaxies and gas-rich dwarfs
lie on structures delineated by normal, high surface brightness galaxies
(Mo et al 1994). 

\section{Follow up observations} 
 The follow up 21cm synthesis observations on the galaxies found by this
survey are essential for a number of reasons.  First, Arecibo is capable
of detecting galaxies as far as 6 arcmin from the center of the beam. 
The positional accuracy is too poor to make unambiguous identifications
with cataloged galaxies or galaxies on the DSS.  Second, reliable flux
measurements are necessary in order to construct an HIMF.  Flux
measurements from the survey data can be poor if the detected galaxy is
more extended than the survey beam or if it is detected at a large
distance from the center of the beam.  Finally, we have found that some
signals were actually caused by pairs or a small group of galaxies. 
This was not always obvious from looking just at survey spectra. 

We took short (20 min) 21cm line observations of 55 of the 61 detected
galaxies with the VLA during the D-Configuration session in May 1995. 
The remaining six galaxies were too close to the sun at the time of the
observations.  These short observations were of sufficient sensitivity
to construct \hi maps and global profiles of 52 of the 55 potential
galaxies.  The signal of three systems fell below the detection limit. 

Optical follow up observations were carried out on the 2.5m Isaac Newton
Telescope on La Palma.  We have been able to make $B$-band images of 24
galaxies during two observing runs in October 1995 and February 1996. 
Additional time has been awarded to observe the remaining galaxies.  So
far, we have been able to make optical identifications of all
\hi selected galaxies.  No isolated \hi clouds without stars have been
found. 

\section{Results}
 The lower panel of Fig.  1 shows the observed distribution of
\hi masses binned per half-decade, with errorbars given by Poison
statistics.  The \hi masses were calculated from either the VLA
observations, or from the Arecibo measurements if the fluxes were lower
than 1.0~Jy~km~s$^{-1}$.  

The inverse of the survey volume as a function
of \hi mass is indicated by the thin line in the upper panel of Fig.  1. 
The curve indicates the upper limit to the space density of
intergalactic \hi clouds without stars as a function of \hi mass. 

The HIMF $\Theta(\mhi)$ was determined following Schmidt's (1968) $\sum
1/V_{\rm max}$ method, which consists of summing the reciprocals
of the volumes corresponding to the maximum distances to which the
objects could be placed and still remain within the sample.  For a
survey such as the Arecibo Strip Survey, $V_{\rm max}$ is a complicated
function, dependent on velocity width, total flux, declination offset
from the center of the survey strip and feed gain, which is a function
of frequency (i.e.  redshift). 

The solid dots in Fig.  1 show the HIMF.  Briggs (1990) derived an
analytical expression for $\Theta$ by using a Schechter luminosity
function and a relation between \hi richness and optical luminosity:
$L$, $\mhi/L \propto L^{\beta}$, where $\beta=-0.1$.  This function is
represented by the fat solid line, using $\mhi*=4.0\times 10^9 h^{-2}
\msol$, a faint end slope $\alpha=1.25$ and a normalization $\theta^* =
0.013$, which is a satisfactory fit to the points.  The parameters of
this fit agree quite well with those of optical luminosity functions. 
Hence, an \hi selected sample of galaxies does not yield a population of
gas rich dwarf galaxies (Tyson and Scalo 1988), that might be missed by
optical surveys.  If a large population of underluminous galaxies does
exist, they must be either \hi deficient, or have extremely low column
densities ($N_{\rm HI}<10^{18} \rm cm^{-2}$).

The cosmological mass density of \hi at the present epoch, $\rhi(z=0)$,
can be determined from the distribution function of \hi mass in
galaxies.  This function is plotted in Fig.  2.  The fat solid line
indicates the converted best fit \hi mass function, the thin line
represents the sensitivity limits.  The distribution function clearly
illustrates that the integral \hi mass density is dominated by high mass
galaxies, $\mhi\approx 10^{9.5} h^{-2} \msol$ which are $L^*$ galaxies. 
From this figure we derive that $\rhi(z=0) = 4.8~\times~10^7\,h
\,M_\odot~\mbox{Mpc}^{-3}$ or $3.3~\times~10^{-33}\,h\,
\mbox{g\,cm}^{-3}$, with a statistical error of $25\%$.  This result
agrees surprisingly well with earlier estimates by Rao and Briggs
(1993), who find the same value by using optically selected galaxies. 
This implies that there is not much neutral gas hidden in objects like
LSB galaxies or intergalactic clouds that are missed by optical surveys. 
The ratio of \hi mass density to the critical mass density of the
universe at $z=0$ is $\ohi(z=0)= (1.8 \pm 0.4) \times 10^{-4}h^{-1}$,
consistent with a smooth decline of $\ohi$ from high $z$ to the present. 

One final result concerns the distribution of central surface
brightnesses.  Since this galaxy sample is selected regardless of any
optical properties, it is well suited to test the distribution function
of optical surface brightnesses.  The hatched area in Fig.  3 indicates
the possible range of values for the distribution function for the 24
galaxies observed so far.  Despite the large variations due to small
number statistics, it is clear that this distribution is consistent with
the `flat' distribution proposed by McGaugh (1996), of which the
boundaries are given by the dashed and dotted line.  It is noteworthy
that no galaxies observed thus far have central surface brightnesses
fainter than $\sim\!24.0\,B$-$\rm mag\, arcsec^{-2}$, even though the
measurement threshold is $\sim\!26.5\,B$-$\rm mag\, arcsec^{-2}$.  We
therefore appear to be observing a lower limit to the central surface
brightness of gas-rich galaxies in the local universe. 

\medskip

\reference Briggs, F.H. 1990, AJ, 100, 999
\reference Briggs, F.H., Rao, S. 1993, ApJ, 417, 494
\reference Briggs, F.H. 1996, these proceedings
\reference Disney, M.J. 1976, Nature, 263, 573
\reference McGaugh, S.S. 1996, MNRAS, 280, 337
\reference Mo, H.J., McGaugh, S.S., Bothun, G.D. 1994, MNRAS, 267, 129
\reference Rao, S., Briggs, F. 1993, ApJ, 419, 515
\reference Schmidt, M.  1968, ApJ, 151, 393
\reference Sorar, E. 1994, Ph.D. Thesis, University of Pittsburgh
\reference Szomoru, A., Guhathakurta, P., van Gorkom, J.H., Knapen,
J.H., Weinberg, D.H., Fruchter, A.S. 1994, AJ, 108, 491
\reference Szomoru, A., van Gorkom, J.H., Gregg, M.D., Strauss, M.A.
1996, AJ, 111, 2150
\reference Tyson, N.D., Scalo, J.M. 1988, ApJ, 329, 618
\reference Weinberg, 1991, ApJ, 372, L13

\begin{figure}[htb] \epsfxsize=350pt \epsfbox{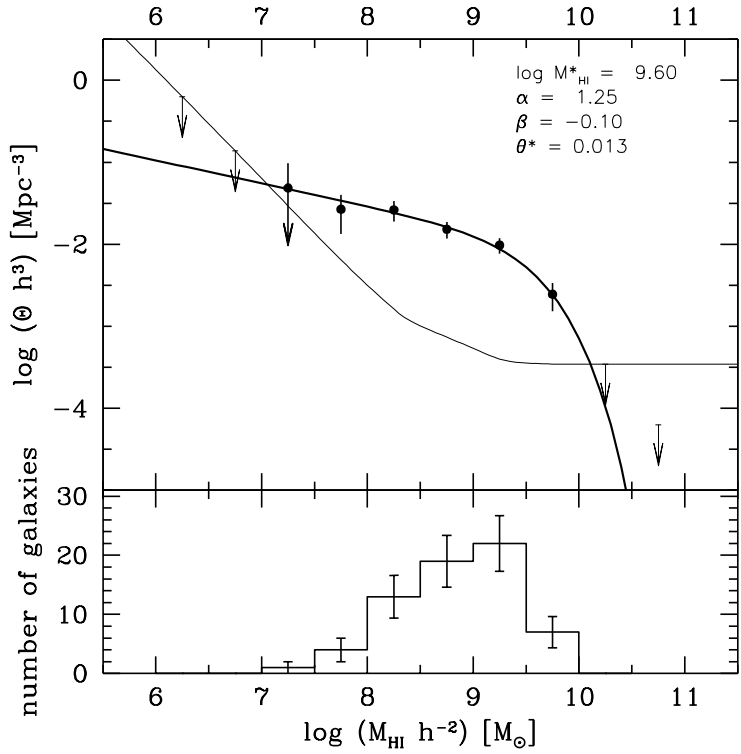} \caption{
Lower panel: The distribution of \hi masses of the detected galaxies,
with errorbars given by Poison statistics. 
 Upper panel: The thin line is the sensitivity of our survey.  The
measured \hi mass function per half decade is shown by the points.  An
analytical HIMF is represented by the fat line, using the parameters
given in the upper right corner.  The arrows show upper limits to the
volume density of \hi clouds.  The two measurements on the right are
from a complementary survey with the Arecibo telescope over the range
19,000 to 28,000~\kms.}
\end{figure}

\begin{figure}[htb]
\epsfxsize=350pt
\epsfbox{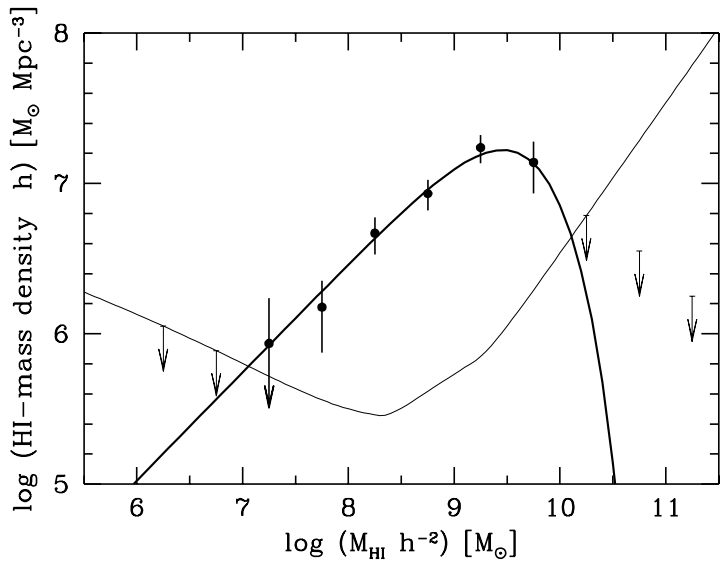}
\caption{ Space density of \hi mass contained in objects of different
masses per half decade. Thin line indicates again the sensitivity of 
the survey.}
\end{figure}

\begin{figure}[htb]
\epsfxsize=350pt
\epsfbox{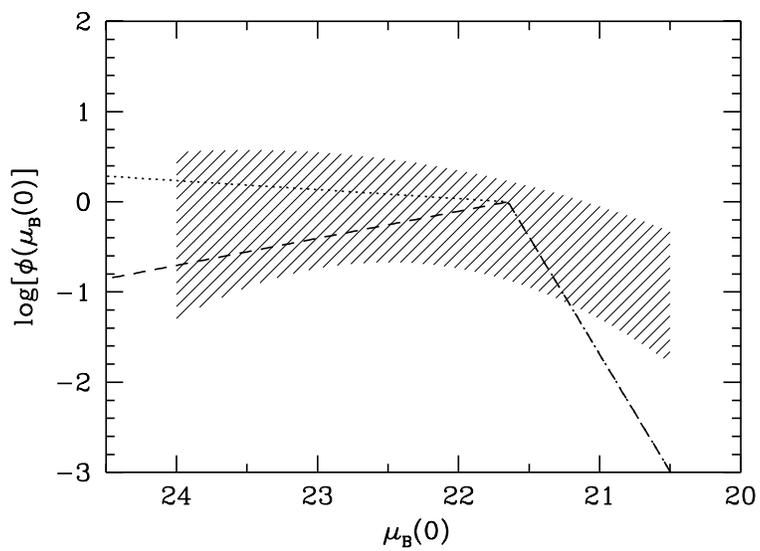}
\caption{ The volume corrected surface brightness distribution of our
\hi selected galaxy sample. Hatched area shows the possible range of
values for this distribution function. The two
lines represent the upper and lower limit to the distribution proposed
by McGaugh (1996).  The y-axis is arbitrarily scaled. }
\end{figure}

\end{document}